# Exploring Scientometrics with the OpenAIRE Graph: Introducing the OpenAIRE Beginner's Kit

Andrea Mannocci[1,*], Miriam Baglioni[2]

[1]*andrea.mannocci@isti.cnr.it*
https://orcid.org/0000-0002-5193-7851
CNR-ISTI, Pisa, Italy

[2] *miriam.baglioni@isti.cnr.it*
https://orcid.org/0000-0002-2273-9004
CNR-ISTI, Pisa, Italy

[*]*corresponding author*

## Abstract

The OpenAIRE Graph is an extensive resource housing diverse information on research products, including literature, datasets, and software, alongside research projects and other scholarly outputs and context. It stands as a cornerstone among contemporary research information databases, offering invaluable insights for scientometric investigations. Despite its wealth of data, its sheer size may initially appear daunting, potentially hindering its widespread adoption. To address this challenge, this paper introduces the OpenAIRE Beginner's Kit, a user-friendly solution providing access to a subset of the OpenAIRE Graph within a sandboxed environment coupled with a Jupyter notebook for analysis. The OpenAIRE Beginner's Kit is meticulously designed to democratise research and data exploration, offering accessibility from standard desktop and laptop setups. Within this paper, we provide a brief overview of the included dataset and offer guidance on leveraging the kit through a selection of illustrative queries tailored to address common scientometric inquiries.

## 1. Introduction

Scientometrics research is traditionally associated with paywalled sources of research information such as Scopus and Web of Science, despite this hampering the full reproducibility and reuse of results and fair access to the discipline.

However, more recently, there has been an ever-growing interest in open research information[1], leading to the emergence of various initiatives potentially fuelling scientometrics research. Examples include OpenAlex (Priem, Piwowar & Orr, 2022), OpenCitations (Peroni & Shotton, 2020), Dimensions (Herzog, Hook & Konkiel, 2020), and the now-defunct Microsoft MAG (Wang, Shen et al., 2020) to name a few.

The OpenAIRE Graph (hereafter the Graph, for brevity) can complement this landscape by providing a thorough perspective on Open Science and parting from a strictly literature-based representation of the global scientific record.

However, housing nearly 240 million research outputs beyond publications, 130,000 data sources, 320,000 organisations, 3 million projects, and over 5 billion relationships between these entities, the Graph size (i.e., about 270 GB compressed JSON files (Manghi, Atzori et al., 2024)), similarly to other research information providers, represents the major significant hurdle for new potential users.

To address this accessibility challenge, for example, OpenAlex and Dimensions facilitated access to their resources by parking data on cloud platforms (e.g., Google Big Query, Amazon Web Services), whose exploitation ultimately falls on users' finances and availabilities in the long run.

---

[1] https://barcelona-declaration.org

Conversely, OpenAIRE developed a Beginner's Kit consisting of a sandboxed environment providing access to a smaller, more manageable subset of the Graph and a notebook showing a collection of example queries against the data. This enhanced accessibility fosters effortless and cost-free initial engagement with the Graph.

The Beginner's Kit not only supports newcomers approaching scientometric data analysis for the first time but also experienced users who are approaching the Graph and want to understand better the types of information it contains and how the data is structured. This acquired knowledge and set of skills is instrumental when researchers begin to work with the whole Graph at a later stage.

Finally, the Beginner's Kit provides a safe environment for programmers to test their code to analyse the Graph data. This allows them to experiment and refine their queries and code on a smaller scale before applying it to complete the Graph, potentially saving valuable time and resources.

## 2. OpenAIRE Graph in a Nutshell

The Graph is one of the most prominent open-access resources for scholarly communication, and it can be a powerful ally in navigating the ever-expanding realm of scientific discovery. The Graph integrates a vast amount of interconnected information (see Fig. 1), such as metadata descriptions about publications, research projects, organisations, and more, as well as the relationships between such entities.

Through the Graph, researchers can explore various facets of the global academic record, uncovering hidden trends and fostering a more comprehensive understanding of the complete research ecosystem.

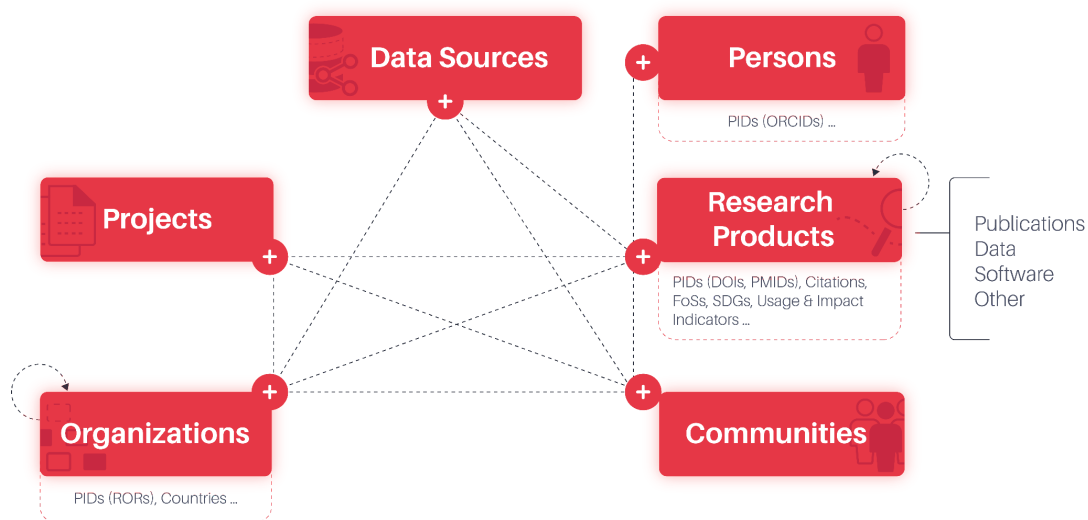

Fig. 1 - Bird's-eye view of the OpenAIRE Graph data model

The foundation of the Graph lies in bibliographic records harvested from reputable sources, such as Crossref, Datacite, ORCID, and many more. This rich core is further integrated with information from registries such as DOAJ, re3data, funders databases, and over 1,000 institutional repositories[2]. The collected metadata are then cleaned (to ensure compliance against chosen vocabularies). Afterwards, several mining algorithms are exploited on the ever-growing set of full texts of open-access publications collected by OpenAIRE (over 25 Million as of March 2024) to enrich the Graph with new relations and properties. The Graph is further enriched with Fields of Science (FoS) and Sustainability Development

---
[2] https://graph.openaire.eu/docs/graph-production-workflow/aggregation

Goals (SDGs) from SciNoBo (Kotitsas, Pappas et al., 2023), usage metrics from Bip!Finder indicators (Vergoulis, Chatzopoulos et al., 2019) and UsageCounts[3], and many more. Then, a meticulous process of disambiguation is undertaken. This step ensures data accuracy by identifying and merging duplicate entities, guaranteeing that each research entity possesses a distinct and unambiguous identifier within the Graph. The overall construction workflow of the Graph is reported in Fig. 2.

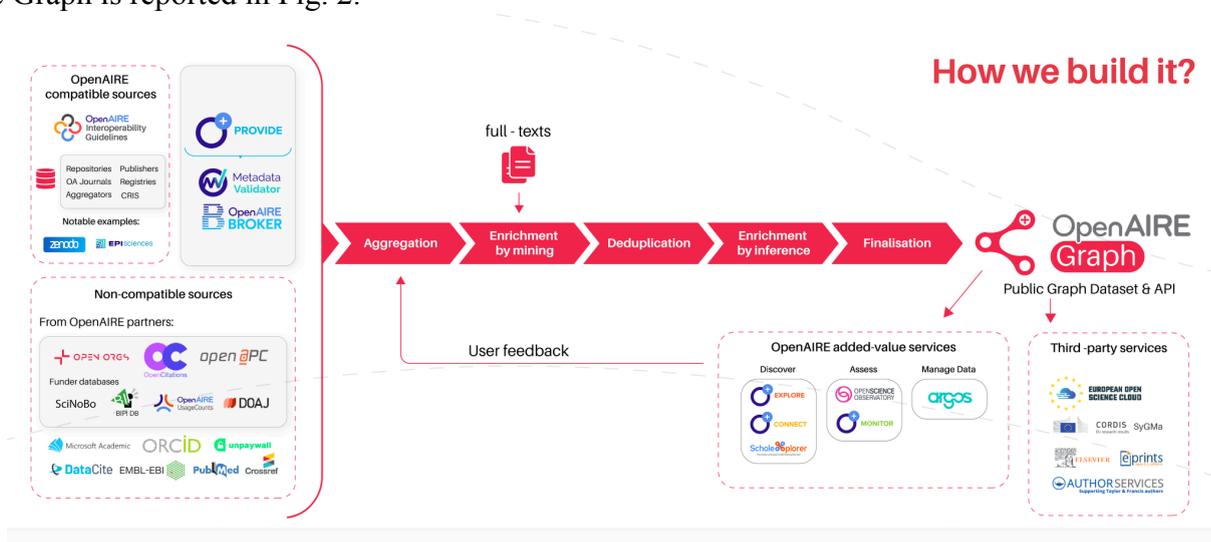

Fig. 2 - Overview of the OpenAIRE constriction workflow

After deduplication, a further enrichment step is performed on the Graph, aiming to further extend the set of relationships between the entities and enrich the properties of the research products. After the finalisation step, the Graph is ready to be shared via APIs and dedicated dumps and exploited by several services.

The Graph construction is fully documented online[4], ensuring transparency of the whole process.

### 3. The OpenAIRE Beginner's Kit

The OpenAIRE Beginner's Kit is available on GitHub[5], and it is deposited on Zenodo as well (Baglioni, Mannocci et al., 2024). It just requires the presence of Docker Desktop[6] on the local machine; fulfilling this requirement suffices for building the Docker image and running locally a Docker container virtualising an Apache Hadoop[7] cluster and providing a Jupyter Lab instance for playing with the data via SparkSQL[8] and Python. The presence of Apache Spark makes it possible to query metadata in JSON files as virtual SQL tables of a relational database and take advantage of data parallelism. In contrast, processing the same data in-memory right away with Python would be unfeasible and incur out-of-memory exceptions.

The Docker image already ships common libraries such as pandas[9] for data manipulation, igraph[10] for network analysis, and matplotlib[11] and plotly[12] for data visualisation. Future releases may include additional tooling and libraries proven useful for common analyses in scientometrics.

---

[3] https://usagecounts.openaire.eu
[4] https://graph.openaire.eu/docs/graph-production-workflow
[5] https://github.com/openaire/beginners-kit
[6] https://www.docker.com/get-started
[7] https://hadoop.apache.org
[8] https://spark.apache.org/sql
[9] https://pandas.pydata.org
[10] https://igraph.org
[11] https://matplotlib.org
[12] https://plotly.com/python

In this section, we describe the main components of the OpenAIRE Beginner's Kit: the dataset in Section 3.1 and the notebook in Section 3.2.

### 3.1. The OpenAIRE Beginner's Kit Dataset

The OpenAIRE Beginner's Kit is intended to be used to learn how to navigate and utilise the whole Graph and how this can be used to address scientometric research questions.

To accomplish this goal, the Beginner's Kit includes a selection of the graph data entities and their semantic relations, providing a practical overview. Since all the entities and relationships must be represented, the Beginner's Kit features research products published up to eight months before the Beginner's Kit date of release (i.e., March 2024, at the time of writing). Starting from this set of research products (Fig. 3.a), all the other kinds of entities reachable following a direct link from any of these research products are included in the kit (Fig. 3.b). Then, all the relationships between the selected entities are also included (Fig. 3.c).

The resulting dataset contains the metadata of research products published between 30 June 2023 and 29 February 2024 for a total of 3,919,148 publications; 808,583 datasets; 27,470 software; 167,367 other research products; 53,546 datasources; 55,633 organisations, 29 communities; 52,675 projects, and 35,315,406 relations.

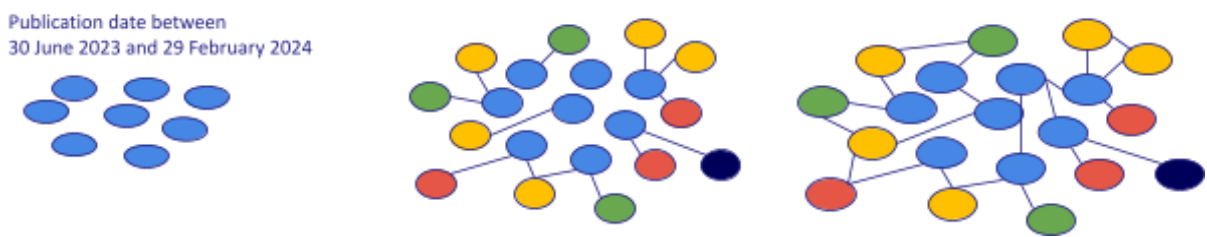

Fig. 3 - Incremental Beginner's Kit Dataset construction

### 3.2. Exploring the Beginner's Kit Notebook

The notebook complementing the dataset in the OpenAIRE Beginner's Kit is organised as a standard Jupyter notebook[13] running in a JupyterLab instance hosted in the Docker container. The notebook is organised into distinct paragraphs; some paragraphs contain just text following the Markdown[14] syntax and function as documentation and guidance, while others contain instead code. Each code paragraph can be run independently, and the output is shown right after once the execution has ended.

Each code paragraph is written in Python, and its core part represents an SQL-like[15] query against the data that replies to a question stated above the block. A number of built-in functions is provided by SparkSQL and documented online[16].

To begin with, the notebook loads the JSON data and prepares several virtual tables following the OpenAIRE schema[17]; every tuple in the Graph has a unique ID valid in OpenAIRE. The (virtualised) tables made available are the following:
- `publications` contains metadata of research products which are intended to disseminate research and read by humans (e.g., article, thesis, peer-review, blog posts, books, reports, patents, etc.);

---

[13] https://jupyter.org
[14] https://www.markdownguide.org
[15] An in-depth overview of differences of the SparkSQL dialect, ANSI and SQL-2016 is documented here, https://spark.apache.org/docs/latest/sql-ref-ansi-compliance.html
[16] https://spark.apache.org/docs/latest/api/sql
[17] https://graph.openaire.eu/docs/data-model

- `datasets` contains metadata for research products referring to self-contained, persistently-identified digital assets intended for processing (e.g., files containing raw data, tables, metadata collections, dumps; persistent dynamic queries to scientific databases);
- `softwares` contains metadata of research products referring to source code files, algorithms, scripts, computational workflows, and executables that were created during the research process or for a research purpose;
- `others` contains any digital asset, uniquely identified, whose nature does not fall under the first three types described above;
- `results` is the union of the four tables above;
- `datasources` contains metadata referring to services where published material (metadata and files) is stored, preserved, and made discoverable and accessible;
- `organizations` contains metadata about academic institutions, research centres, funders, or any other institutions taking part in the research process;
- `projects` contains metadata describing funding awarded to a person or an organisation by a funding body;
- `communities` contains metadata about the research communities/infrastructures registered in OpenAIRE;
- `relations` contains the relations between couples of source and target IDs. The semantics of the relations follow the DataCite semantic definition, extended by specific semantics[18].

For example, counting the number of citations accrued by publications is as simple as joining the table `publications` with the table `relations`, filtering by relation type to select the proper citational semantics (i.e., `IsCitedBy`), grouping by publication identifier and counting. The resulting query is as follows.

```sql
SELECT publications.id, pid.value, COUNT(*) AS count
FROM publications
JOIN relations
  ON publications.id = relations.source
WHERE reltype.name = 'IsCitedBy'
GROUP BY publications.id, pid.value
ORDER BY count DESC
```

A more complex example could be computing the breakdown of research product access right categories aggregated by organisations' countries. This is achieved by leveraging the `isAuthorInstitutionOf` relationship. This relation links organisations (source) to research products where at least one author is affiliated with that organisation. It is important to note that this affiliation is directly between products and organisations, not between authors and organisations (i.e., it does not represent an authorship linkage). Once this initial connection is established, the second step focuses on linking affiliated organisations with the access rights associated with their corresponding research products. By grouping data based on the organisation's country code, we can calculate the prevalence of different access rights (e.g., open access) within each geographical region. The `COUNT(IF(bestaccessright.label = 'OPEN', 1, NULL)` operation serves as a counting tool, specifically tallying the occurrences of the desired access type (open access in this case) based on the specified condition. This approach can be readily adapted to analyse the prevalence of other access right categories in the query.

---

[18] https://graph.openaire.eu/docs/data-model/relationships/relationship-types

```sql
SELECT organizations.country.code AS country,
  COUNT(*) AS total,
  COUNT(IF(bestaccessright.label = 'OPEN', 1, NULL)) AS open,
  COUNT(IF(bestaccessright.label = 'EMBARGO', 1, NULL)) AS embargo,
  COUNT(IF(bestaccessright.label = 'CLOSED', 1, NULL)) AS closed
FROM organizations
JOIN relations
JOIN results
  ON organizations.id = relations.source AND
  results.id = relations.target AND
  reltype.name = 'isAuthorInstitutionOf'
WHERE organizations.country IS NOT NULL
GROUP BY organizations.country.code
ORDER BY total DESC
```

The same query type can also be exploited to investigate the open access publishing practices across organisations by leveraging the organisation's legal name as a query parameter rather than the organisation's country. This refined approach offers a more granular view of open access adoptions within specific institutions. Furthermore, for a deeper understanding of open access dynamics, we can incorporate the publication year into the query equation (see the query below). This additional dimension empowers researchers to analyse the timewise evolution of open access publishing practices.

```sql
SELECT COALESCE(legalshortname, legalname) AS organization,
  COUNT(*) AS total,
  year(publicationdate) AS pub_year,
  COUNT(IF(bestaccessright.label='OPEN', 1, NULL)) AS open,
  COUNT(IF(bestaccessright.label='EMBARGO', 1, NULL)) AS embargo,
  COUNT(IF(bestaccessright.label='CLOSED', 1, NULL)) AS closed
FROM organizations
JOIN relations
JOIN results
  ON organizations.id = relations.source
  AND results.id = relations.target
  AND reltype.name = 'isAuthorInstitutionOf'
GROUP BY organization, pub_year
ORDER BY organization, total DESC
```

The results of a query, as long as they can fit into memory, can be passed to other libraries as well for further analysis. For example, the following query isolates couples of countries co-participating in project grants and the relative number of projects they co-participate in. The results of the query can be, for example, passed to igraph as a weighted edge list to load the network for further analysis and visualisation, as represented in Fig. 4. Similarly, data could be extracted via queries, exported in the proper format and then loaded onto VOSviewer (Van Eck, & Waltman, 2010).

```
WITH countryProject AS (
  SELECT country.code AS country, target AS id
  FROM organizations
  JOIN relations
    ON reltype.name = 'isParticipant' AND
    source = organizations.id
  WHERE country IS NOT NULL )
SELECT l.country AS left, r.country AS right, COUNT(*) AS count
FROM countryProject AS l
JOIN countryProject AS r
  ON l.id = r.id AND l.country <= r.country
GROUP BY left, right
ORDER BY count DESC
```

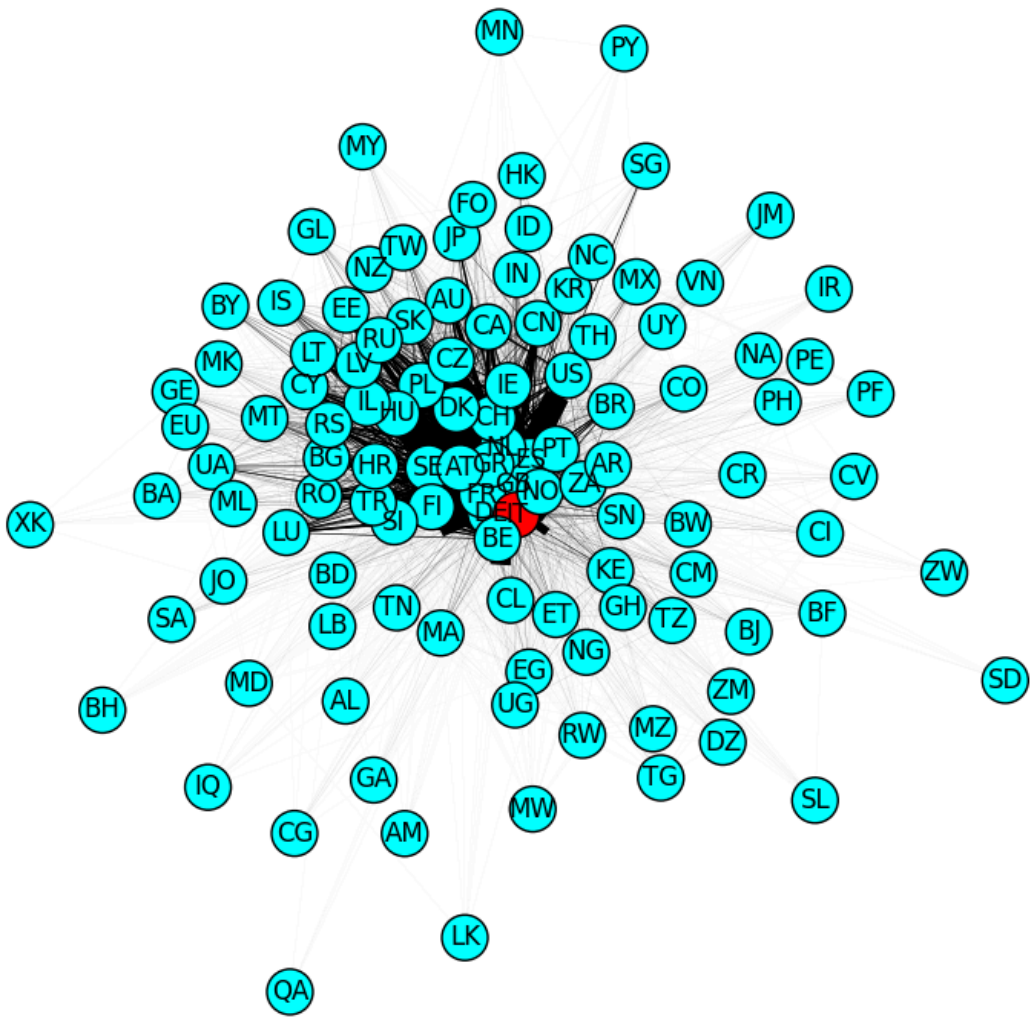

Fig. 4 - Network of countries co-participating in research grants with Italy. Node labels represent the country code, while edge thickness encodes the normalised edge weights (i.e., the number of projects a couple of countries co-participate in).

## 4. Current limitations and future works

In the future, we plan to extend the OpenAIRE Beginner's Kit with respect to both the dataset and the notebook.

For the time being, the provided dataset prevents meaningful diachronic analyses, as the included research products are drawn from the last eight months of the global research record, spanning partially over 2023 and 2024. To tackle this aspect, we will devise better subsetting strategies and resample the Graph in order to fetch a representative population of research products spanning multiple years and provide adequate coverage across disciplines and organisations.

Likewise, the notebook will be extended with further queries and possibly reorganised by grouping them into multiple focused notebooks, each targeting a particular facet of scientometric inquiry (e.g., citation analysis, spatial scientometrics, collaborations). Additionally, separate notebooks will be tailored for funders and organisations. These stakeholder-specific notebooks will consider their specific needs by investigating scientometric aspects most relevant to them.

In conclusion, we nonetheless reckon that sandboxed environments such as the OpenAIRE Beginner's Kit can be an invaluable aid, and we believe that these should flank data availability on the cloud to democratise access and first-entry experimentation of large datasets for scientometrics analysis.

**Open science practices**

The key components of the OpenAIRE Beginner's Kit, the notebook and the dataset, are both available online on GitHub and Zenodo for long-term preservation and formal citation purposes. Future releases of the OpenAIRE Beginner's Kit will be deposited automatically on Zenodo thanks to the integration present between the two systems.

**Author contributions**

Andrea and Miriam equally contributed to this publication (https://credit.niso.org/contributor-roles/writing-original-draft).

**Competing interests**

There are no competing interests to declare.

**References**

Baglioni M., Mannocci A., Bloisi, G., & La Bruzzo S. (2024). openaire/beginners-kit: OpenAIRE Graph Beginners Kit v3.1 (v3.1). Zenodo. https://doi.org/10.5281/zenodo.10843122

Herzog, C., Hook, D., & Konkiel, S. (2020). Dimensions: Bringing down barriers between scientometricians and data. Quantitative Science Studies, 1(1), 387–395. https://doi.org/10.1162/qss_a_00020

Kotitsas, S., Pappas, D., Manola, N., & Papageorgiou, H. (2023). SCINOBO: A novel system classifying scholarly communication in a dynamically constructed hierarchical Field-of-Science taxonomy. Frontiers in Research Metrics and Analytics, 8. https://doi.org/10.3389/frma.2023.1149834

Manghi, P., Atzori, C., Bardi, A., Baglioni, M., Dimitropoulos, H., La Bruzzo, S., Foufoulas, I., Mannocci, A., Horst, M., Iatropoulou, K., Kokogiannaki, A., De Bonis, M., Artini, M., Lempesis, A., Ioannidis, A., Manola, N., Principe, P., Vergoulis, T., Chatzopoulos, S., & Pierrakos, D. (2024). OpenAIRE Graph Dataset (7.0.0). Zenodo. https://doi.org/10.5281/zenodo.10488385

Peroni, S., & Shotton, D. (2020). OpenCitations, an infrastructure organization for open scholarship. Quantitative Science Studies, 1(1), 428–444. https://doi.org/10.1162/qss_a_00023

Priem, J., Piwowar, H., & Orr, R. (2022). OpenAlex: A fully-open index of scholarly works, authors, venues, institutions, and concepts (arXiv:2205.01833). arXiv. https://doi.org/10.48550/arXiv.2205.01833

Van Eck, N. J., & Waltman, L. (2010). Software survey: VOSviewer, a computer program for bibliometric mapping. Scientometrics, 84(2), 523–538. https://doi.org/10.1007/s11192-009-0146-3

Vergoulis, T., Chatzopoulos, S., Kanellos, I., Deligiannis, P., Tryfonopoulos, C., & Dalamagas, T. (2019). BIP! Finder: Facilitating Scientific Literature Search by Exploiting Impact-Based Ranking. Proceedings of the 28th ACM International Conference on Information and Knowledge Management - CIKM '19, 2937–2940. https://doi.org/10.1145/3357384.3357850

Wang, K., Shen, Z., Huang, C., Wu, C.-H., Dong, Y., & Kanakia, A. (2020). Microsoft Academic Graph: When experts are not enough. Quantitative Science Studies, 1(1), 396–413. https://doi.org/10.1162/qss_a_00021